\documentclass[a4paper,12pt]{article}

\usepackage{float}
\usepackage{a4wide}
\usepackage{fullpage}
\usepackage[bf,sf]{titlesec}
\usepackage[margin=27pt,font=small,labelfont=bf]{caption}
\usepackage{fancyhdr}
\usepackage{amsmath,amssymb}
\usepackage{graphicx}

\newcommand{\mytitle}{\large Human sperm cells swimming in micro-channels}
\newcommand{\myauthors}{\normalsize P. Denissenko$^{15*}$, V. Kantsler$^{2}$, D.J. Smith$^{315}$ and J. Kirkman-Brown$^{45}$}

\begin{document}

\title{\textbf{\mytitle}}
\author{\myauthors}
\date{}


\title{\textbf{\textsf{Human sperm cells swimming in micro-channels}}}

\renewcommand*\abstractname{\textsf{Abstract}}

\newcommand{\bs}{\boldsymbol}
\newcommand{\smum}{\, \mu \mathrm{m}}
\newcommand{\mums}{\, \mu \mathrm{m}/\mathrm{s}}

\maketitle
\begin{tabular}{ll}
~~$^1$ &School of Engineering, University of Warwick, Coventry, CV4 7AL, UK.\\[1mm]
~~$^2$ &Department of Applied Mathematics and Theoretical Physics,\\
       &University of Cambridge, Cambridge, CB3 0WA, UK. \\[1mm]
~~$^3$ &School of Mathematics, University of Birmingham, Edgbaston, Birmingham,\\
       &B15 2TT, UK.\\[1mm]
~~$^4$ &School of Clinical and Experimental Medicine, University of Birmingham,\\
       &Edgbaston, Birmingham, B15 2TT UK. \\[1mm]
~~$^5$ &Centre for Human Reproductive Science, Birmingham Women's NHS \\
       &Foundation Trust, Mindelsohn Way, Birmingham, B15 2TG, UK. \\[1mm]
~~$^*$~& Author for correspondence, p.denissenko@warwick.ac.uk
\end{tabular}

\vspace*{\fill} \abstract{The  migratory  abilities  of  motile  human  spermatozoa in vivo are essential for
natural fertility, but it remains a mystery what properties distinguish the tens of cells which find an egg from the millions of cells ejaculated. To reach the site of fertilization, sperm must traverse narrow and convoluted channels, filled with viscous fluids. To elucidate individual and group behaviors that may occur in the complex three-dimensional female tract environment, we examine the behavior of migrating sperm in  assorted  micro-channel geometries. Cells rarely
swim  in  the central  part of  the  channel  cross-section,  instead traveling along the intersection of the channel
walls (`channel corners'). When the channel turns sharply, cells leave  the corner,  continuing  ahead until hitting
the  opposite wall  of the  channel,  with a distribution of departure angles,  the  latter  being modulated by fluid
viscosity. If the  channel bend  is smooth,  cells depart from the inner wall when  the  curvature radius is less than
a threshold value close to 150~$\mu$m. Specific wall shapes are able to preferentially direct motile cells. As a
consequence of swimming along the corners, the  domain occupied  by cells  becomes  essentially 1-dimensional. This
leads to frequent collisions and needs to be accounted for when modeling the behavior of populations of migratory
cells and considering how sperm populate and navigate the female tract. The combined effect of viscosity and three-dimensional architecture should be accounted for in future in vitro studies of sperm chemoattraction.} \vspace*{\fill}
\newpage

\section*{Introduction}
Sperm motility is influenced by surfaces; this is most simply and strikingly evident in the accumulation of cells on
the surfaces of microscope slides and coverslips, a phenomenon known to every andrologist. The effect and its causes
have been investigated extensively through a variety approaches, including microscopy
\cite{rothschild1963,winet1984,cosson2003,woolley2003}, computational fluid mechanics,
\cite{ramia1993,fauci1995,smith2009jfm,smith2009ms,blake1971}, molecular dynamics \cite{elgeti2010} and mathematical
analysis \cite{crowdy2010pre}. Principal points addressed by previous studies are the extent to which surface
accumulation is a generic feature fluid dynamic effect associated with near-wall swimming, the role of specialized
flagellar beat patterns, species-specific morphology, and the relative prevalence of swimming `near' as opposed to
`against' walls; discussion of these questions can be found in recent editorials \cite{smith2011bj,elgeti2011}. There
has also been a resurgence of interest recently in the fluid mechanics of motile bacteria
\cite{lauga2006,li2009,giacche2010pre,shum2010}, and generic models for swimming cells
\cite{berke2008,or2009,crowdy2010jfm,crowdy2010pre}.

Previous studies have usually focused on the behavior of a cell near a single planar surface or between a pair of
planar surfaces, modeling the interior of a haemocytometer or similar device; however, both the female reproductive
tract and microfluidic {\it in-vitro} fertilization (IVF) devices present sperm with a much more confined and
potentially tortuous geometry. The fallopian tubes consist of ciliated epithelium \cite{suarez2006}, the distance
between opposed epithelial surfaces being of the order of $100$~$\mu$m in many regions, particularly cervical crypts
and the folds of the ampullary fallopian tube, comparable with the approximate $50$~$\mu$m length of the human sperm
flagellum. Microchannel fabrication technology also allows the construction of environments with complex geometries
that may be exploited in directing and sorting cells \cite{hulme2008}. In this letter we report experimental
observations of the motility of populations of human sperm in fabricated microchannel environments, and the effect of
fluid viscosity.

Bacterial cell movement in microchannels, particularly those produced with soft lithography, has perhaps received more
attention than sperm, and studies have focused more closely on cell tracking and motility characteristics in the
channels. Galajda et al.\ \cite{galajda2007} showed that a `wall of funnels' can be used to concentrate bacteria
preferentially on one side, producing a non-uniform distribution from an initially uniform one --- an apparent example
of `Maxwell's Demon'. Hulme et al.\ \cite{hulme2008} showed that a `ratchet' geometry microchannel can be used to
direct bacterial movement,and that cells can be sorted by length through their ability to navigate different curvature
bends, purely on the basis of cell motility and surface interaction; no external pumping was required. Recently, Binz
et al.\ \cite{binz2010} investigated the effect of channel width and path tortuosity on \textit{S. marcescens}
migration in PDMS microenvironments. These studies lead us to ask --- {\it what principles govern sperm motility in
microchannel environments, how might they be exploited in IVF technology, and how might they extrapolate to
understanding the migration of sperm to the egg?}



\section*{Results}
\begin{figure}[H]
 \vspace{-15mm}
 \hspace{10mm}\includegraphics[width=150mm]{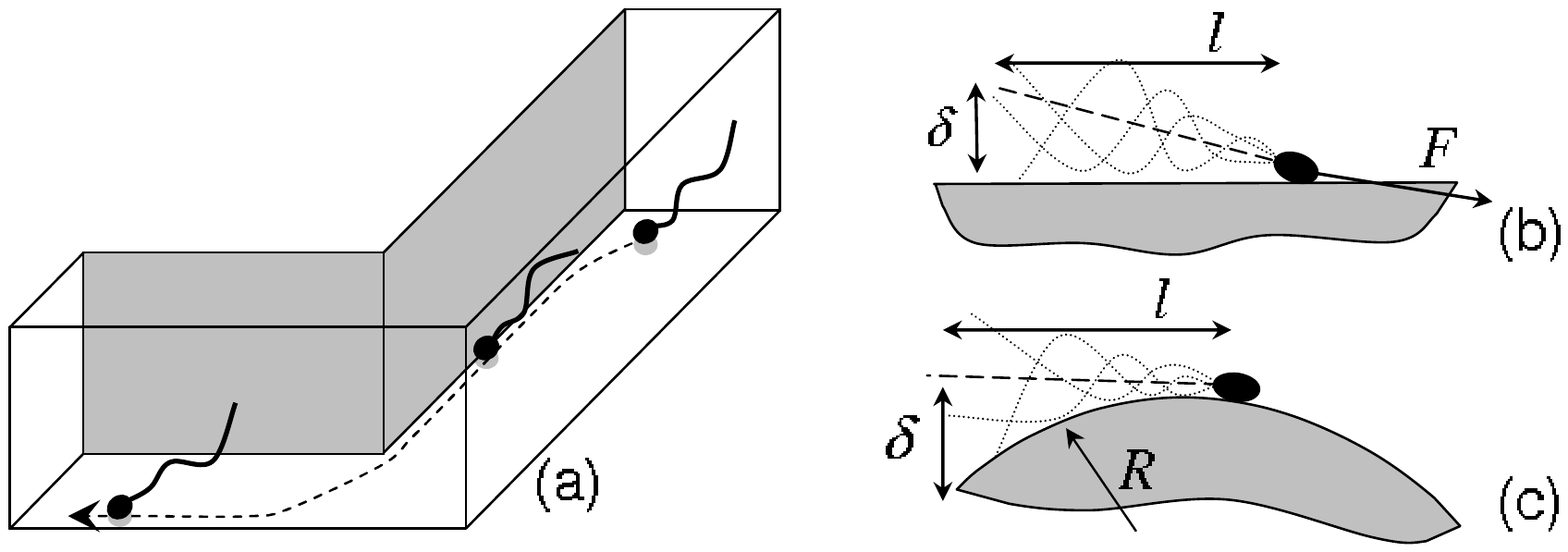}
 \vspace{-155mm}
 \caption{\small Schematic of inferred cell migratory behavior. Cells swim head against the wall, ending up swimming along corners; On sharp turns, cells depart from
channel walls (a). Qualitative explanation of why the cell swim head against the wall (b) and an estimate of the cell
minimum turning radius (c)}\label{sperm:swim_schematics_abc}
\end{figure}

\begin{figure}[H]
 \vspace{-180mm}
 \hspace{10mm}
 \includegraphics[width=180mm]{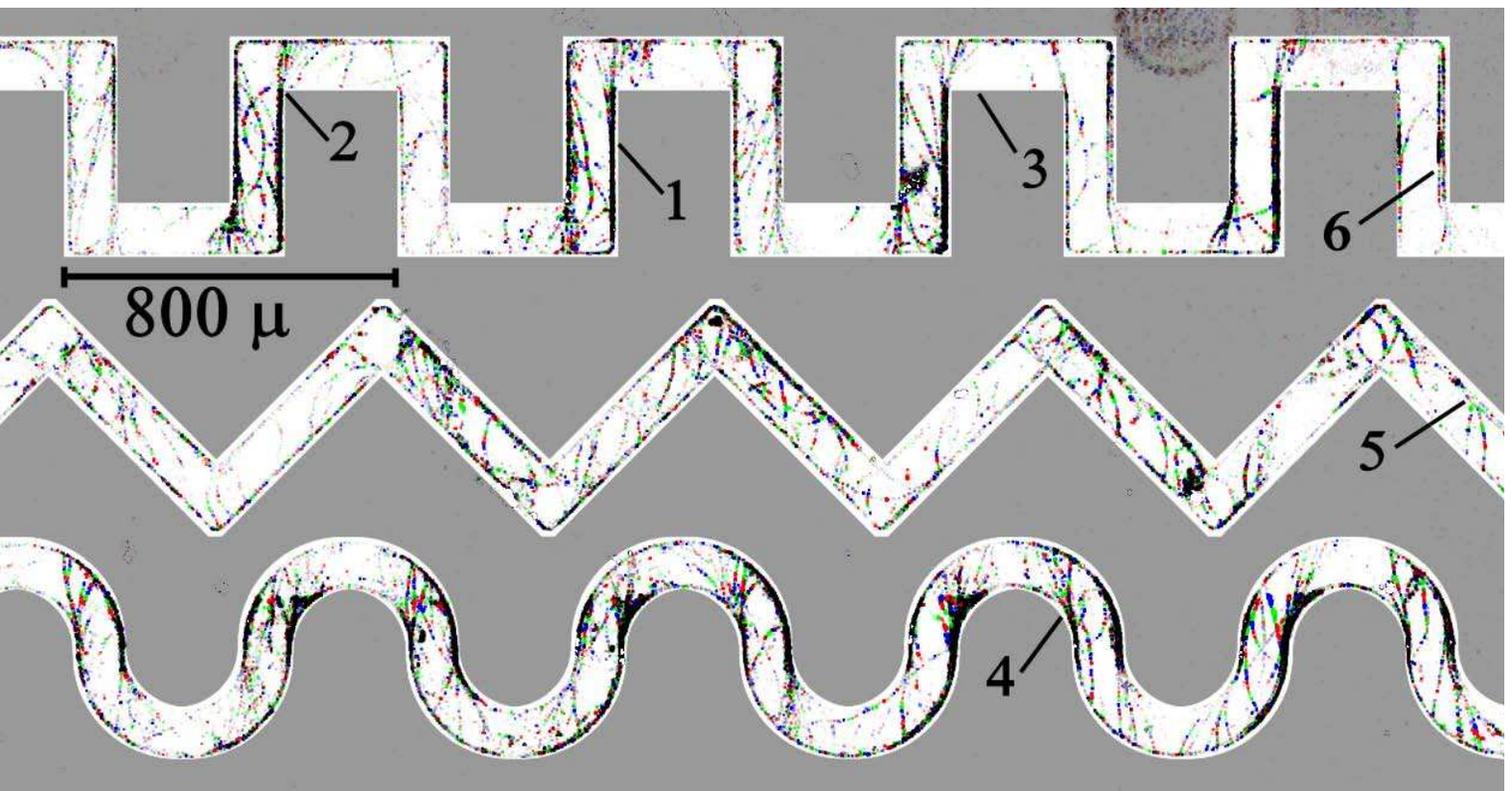}
 \vspace{-5mm}
\caption{\small A typical superposition of an image sequence; top view of the micro-channel. Cell positions in successive frames are color-coded as RGB to resolve the swimming direction. The space between micro-channels is shaded gray to indicate position of walls. Edges of grey shading are spaced from channel walls by approximately 15 micron so that they do not interfere with tracks of the cells. Most of cells swim along the intersection of the channel vertical and horizontal walls (1) with few tracks observed in the middle of the channel. At the periphery of the image where the `side' wall of the channel is observed at an angle, cells traveling along in top and bottom corners between channel walls can be distinguished (6). When the channel turns, cells depart from the wall (2). As a result, no cells travel along the inner corners after the turn (3). In a curved channel, some cells continue to travel along the wall and some depart (4). Cells may also depart from the wall on collision with each other (5) which is shown in Fig.~\ref{sperm:collisions} with a greater magnification.}\label{sperm:image_channel_shapes}
\end{figure}

The first observation is that cells mainly swim along the channel corners as sketched in
Fig.~\ref{sperm:swim_schematics_abc}a. Indeed, contours of the channel appear black (Label~1 in
Fig.~\ref{sperm:image_channel_shapes}), which indicates that many cells passed during the imaging period, as red,
green, and blue stains combine to give black. At the periphery of the frame, due to the short distance between
objective and the channel, the vertical channel wall is visible, enabling us to distinguish cells swimming in the `top'
and `bottom' corners of the channel: We see two parallel bunches of cell tracks indicated by Label~6. Swimming can be
characterized as being almost \textit{against} rather than simply \textit{near} walls, similar to chinchilla sperm
observations of Woolley \cite{woolley2003}, and differing from the mixture of near- and against-wall swimming evident
from experiment \cite{winet1984} in 400~$\mu$m capillary tubes, and computation \cite{smith2009jfm}. This disparity may
be due to the presence of vertical in addition to horizontal walls, and emphasizes the difference between motility in
standard (broad) in vitro environments, where vertical walls are usually not an immediate influencing factor and hence
the cells traverse a 2D wall, as opposed to confined spaces of artificial microchannels and female tract physiology
where the cell will experience a complex 3D series of surfaces.

\begin{figure}[H]
 \vspace{-90mm}
 \hspace{50mm}\includegraphics[width=100mm]{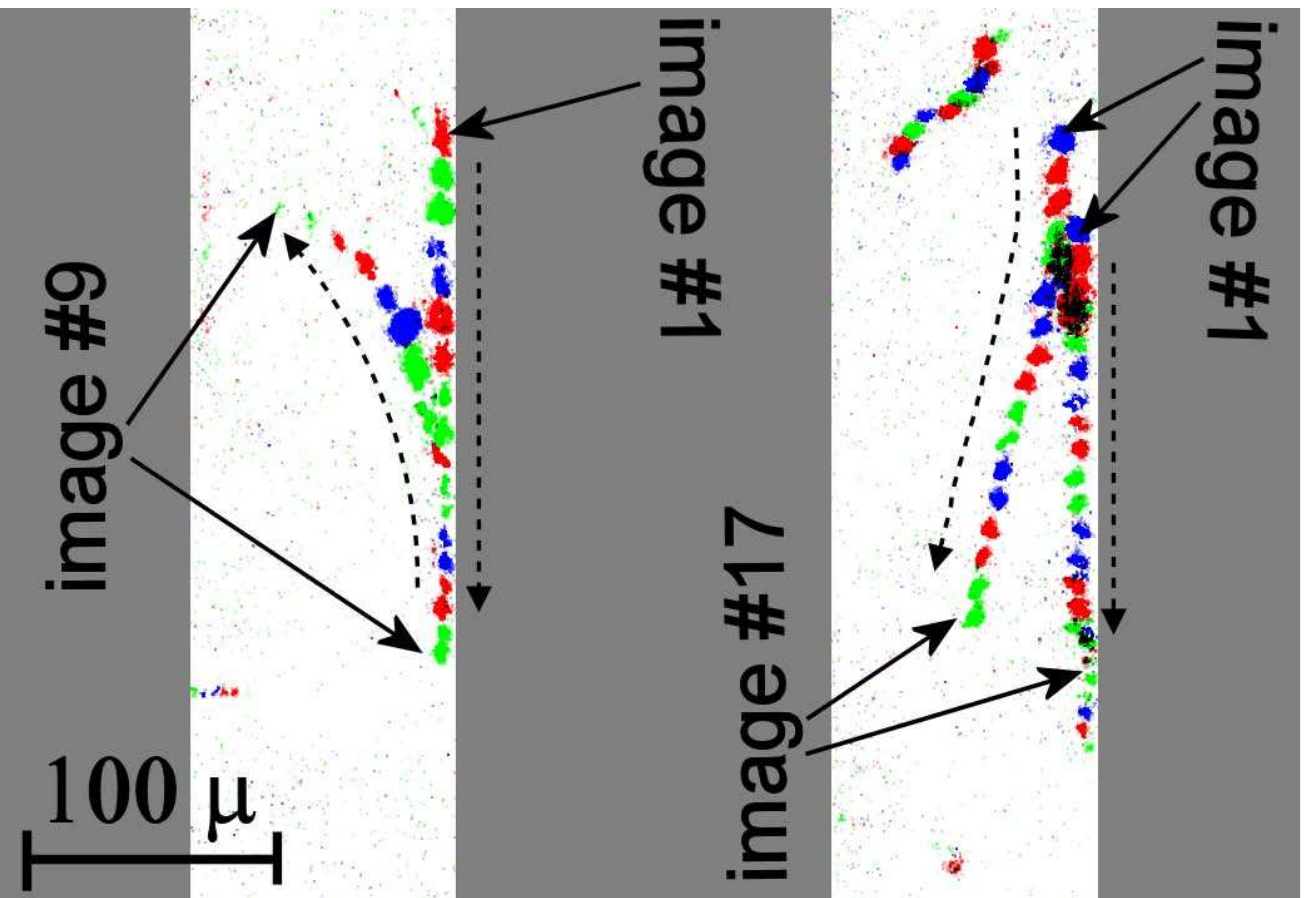}
\caption{\small Cells may depart from walls on collision. The image on the left is composed of 9 consequent frames and
shows a head-on collision; here, the beginning of the track of the departed cell is overdrawn by the track of the cell
which stayed attached and is not visible. The image on the right is composed of 17 consequent frames and shows a
collision when one sperm cell overtakes another. The time interval between images is 1/4 s. Cell swimming directions
are indicated with dotted arrows, positions of the cells in first and last images of sequences are indicated by solid
arrows. Location of the channel walls are indicated by grey shading.}\label{sperm:collisions}
\end{figure}

The next clearly observed effect is that cells depart from walls on sharp turns forming `fans' of trajectories, shown
by Label~2. After reaching the opposite wall, most cells follow it to the next turn. As a result, few or no cells swim
along `inner' segments of channel walls (Label~3). On curved turns, cells may also depart the channel wall (Label~4)
though some cells still continue following the wall. Sometimes cells leave the corner in the absence of geometrical
features (Label~5) which we attribute to collisions. These collisions may be head-on or on overtaking, as shown in
Fig.~\ref{sperm:collisions}.

\begin{figure}[H]
 \includegraphics[width=160mm]{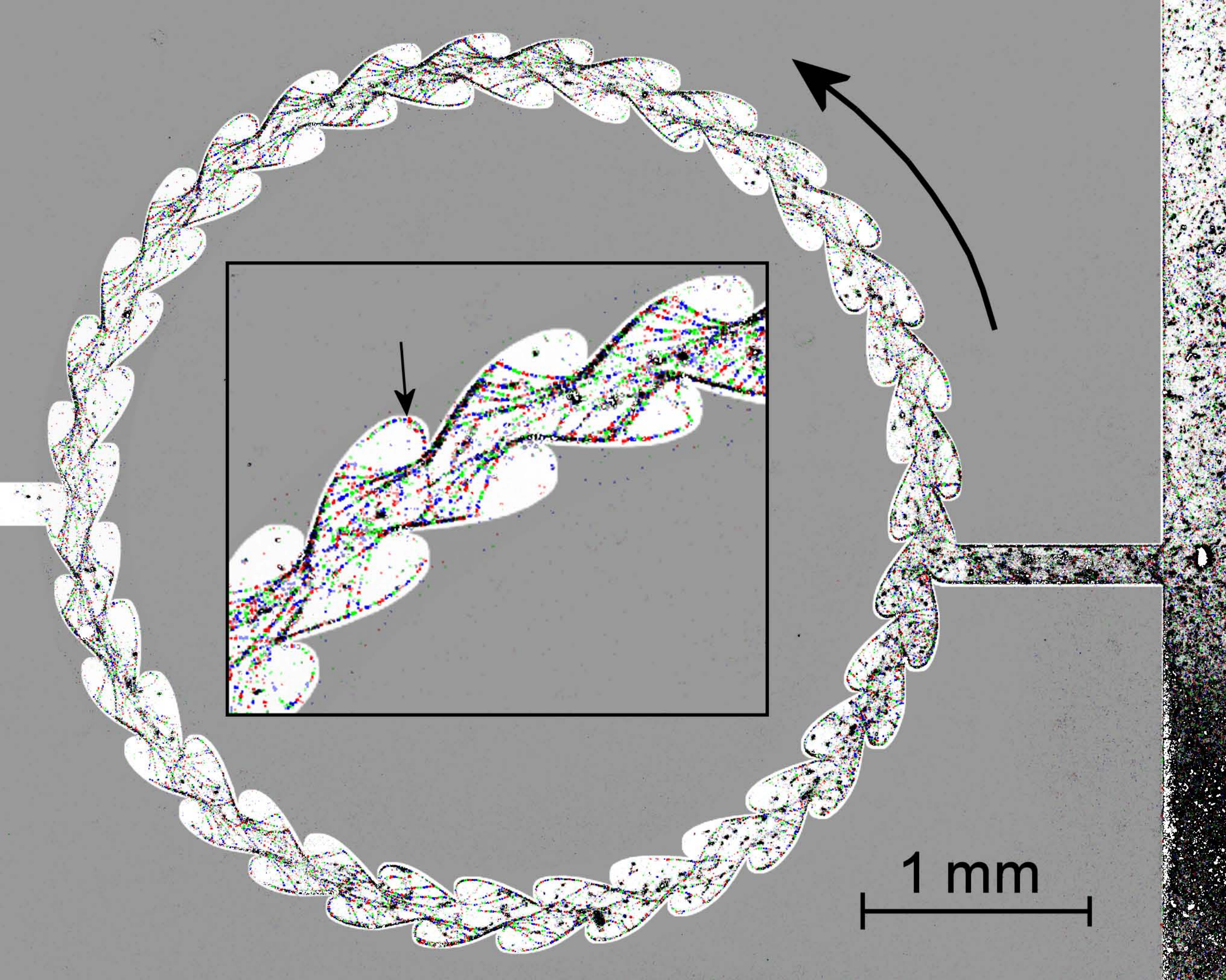}
\caption{\small Spermatozoa in the `one way running track' microchannel geometry. The space outside the micro-channel is shaded gray to indicate position of the walls. Edges of grey shading are spaced from channel walls by approximately 10-20 micron so that they do not interfere with tracks of the cells. The long arrow shows the preferred
(counterclockwise) direction of cell migration. The arrow in the zoomed insert of the channel segment points at a track
of a cell swimming in the direction opposite to that dictated by features of channel walls. Follow the track to see
this cell departing from the inside of the ratchet and traversing the channel, being re-directed anticlockwise, as the
other cells travel.}\label{sperm:one_way_running_track}
\end{figure}

The fact that cells depart from corners can be used to create a channel with ratchet-type walls to force cells to swim
in one direction. Cells in a sort of a circular running track are shown in Fig.~\ref{sperm:one_way_running_track}.
Certain configurations lead to entrapment of cells for extended times. A defective link in an earlier version of a
channel was able to trap cells for as long as 10 minutes before they escape: two crypts on the opposite walls were
staggered in such a way that, while following the channel wall, a cell was ejected by one crypt to get into the other
and then ejected by the latter to return to the first crypt.

\begin{figure}[H]
 \vspace{-30mm} \hspace{25mm} \includegraphics[width=100mm]{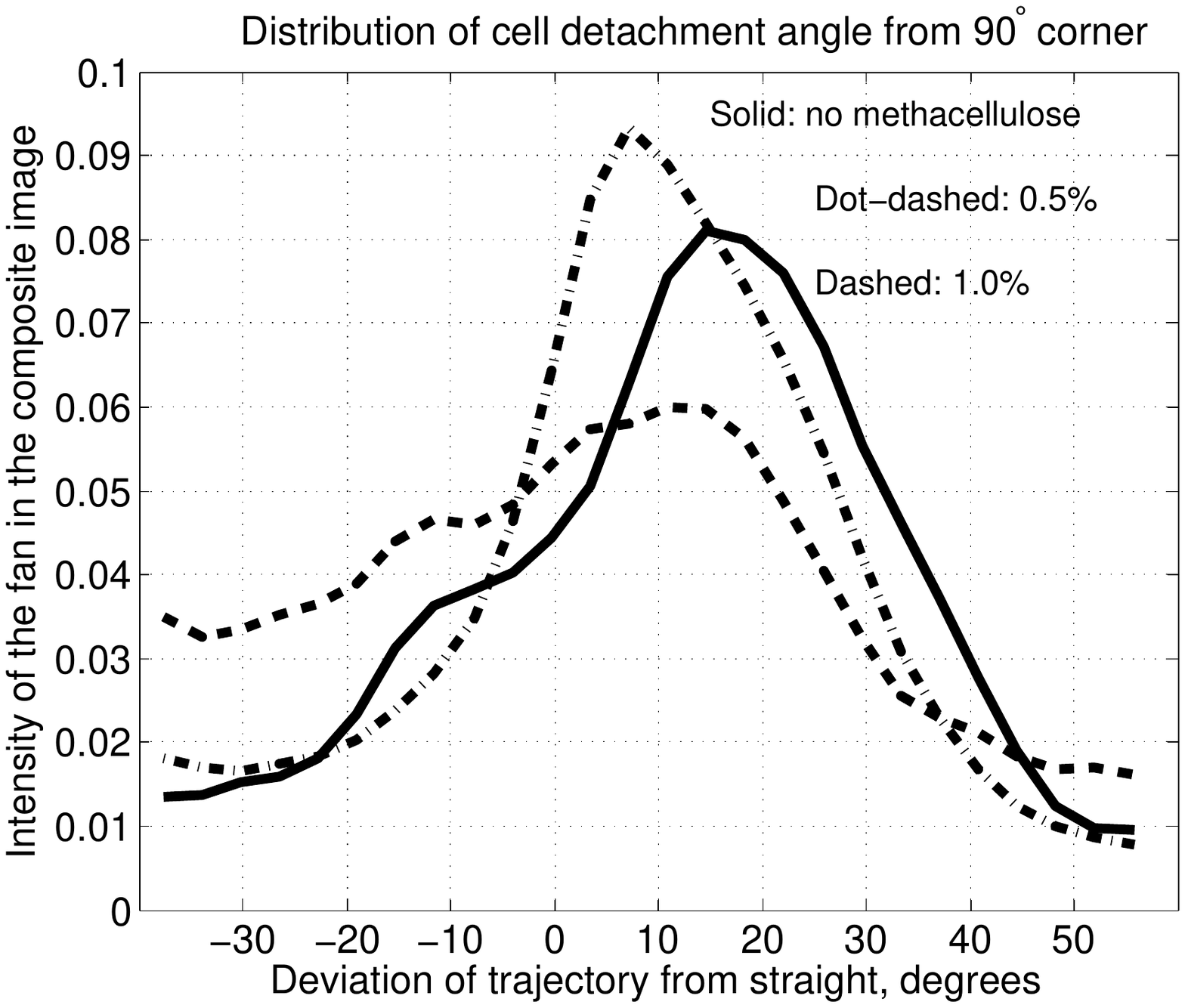}
 \vspace{-30mm} \caption{\small Distribution of the angle of cell departure from the inner wall on a 90$^\circ$ bend of the
channel. Zero angle corresponds to the cell continuing motion without turning, positive angles correspond to cells
turning in the same direction as the channel bends.}\label{sperm:detachment_distributions}
\end{figure}

We have studied the influence of medium rheology on the cell near-wall behavior by filling the microchannel with $0\%$,
$0.5\%$, and $1.0\%$ solutions of methylcellulose. The main effects are shown to be robust with respect to medium
rheology: spermatozoa swim head-against-the-wall and depart from sharp bends in both pure (Newtonian) medium and in the
medium with methylcellulose, which has more than 100 times higher viscosity and complicated rheological properties. A
qualitative observation is that at higher concentration of methylcellulose, visibly more cells swim in the middle of
the channel. To assess the distribution of cells departing from walls on the channel bends, we analyze the pixel
intensity in fans of trajectories starting from channel bends in superposition of image sequences. As the light
sensitivity of our CCD camera is linear to a good approximation, pixel intensity is a suitable quantitative parameter
to use for reconstruction of the cell distribution by departure angles. The 30 minutes long records have been analyzed
and data over four 90$^\circ$ channel bends have been analyzed. Typical results are shown in
Fig.~\ref{sperm:detachment_distributions}b. Depending on the donor, the mean cell turning angle varies from 10$^\circ$
to 20$^\circ$ with the width at half maximum at the level of 25$^\circ$. Observe that a notable part of cells turn away
from the wall (negative angles). No consistent dependence of the departure angle on the concentration of
methylcellulose has been detected, with the concentration affecting the sperm from different donors in different ways.
This can be attributed to a sophisticated interplay between the flagellum stroke pattern and the medium rheological
properties.

\section*{Discussion}
We can interpret our observations
through the following intuitive model, similar to that advanced by Woolley \cite{woolley2003}. The amplitude of head
oscillations is less that that of the end of the tail, so the head can, on average, be closer to the wall. The conical
envelope of the flagellar wave aligns with the surface, resulting in the direction of propulsion being inclined towards
the wall (Fig.~\ref{sperm:swim_schematics_abc}b). Cells therefore are directed towards surfaces, and moreover cells
stay against those surfaces. Once a cell reaches a horizontal wall, it is likely to travel along horizontally while
translating until it, by chance, reaches a vertical wall (or vice versa). It then remains trapped by both walls,
swimming along their intersection, until it finally reaches a sufficiently sharp change to the curvature of a vertical
wall (Fig.~\ref{sperm:swim_schematics_abc}a) to cause departure.


Furthermore, we can use this intuitive model to estimate the minimum turning radius of the cell. Consider similar
triangles, the one formed by the cell envelope and the one formed by radii connecting head and tail of the cell with
the center of a circle forming the channel wall (Fig.~\ref{sperm:swim_schematics_abc}c). Equating the ratios of
triangle bases and sides, we get that the radius of the channel wall $R$ at which a cell of the length $\ell$ is
oriented tangent to the surface at the point of the head contact can be estimated as
\begin{eqnarray}
 &\frac{\delta}{\ell}\approx\frac{\ell/2}{R}\,\,\,\Rightarrow\,\,\,R\approx\frac{\ell}{2\delta}\ell\label{sperm:maxcurvature}
\end{eqnarray}
Substituting $\ell = 50~\mu$m and $\ell/2\delta\sim 3$ from microscopic observations, we get $R\sim$ 150~$\mu$m. This
value is close to the inner radius of curved (the lowest) channel in Fig.~\ref{sperm:image_channel_shapes}. Observe
that while most cells depart on the turn, some stay at the wall which is an indication that the wall radius is not far
from critical in accordance with the estimate (\ref{sperm:maxcurvature}).

The effect of viscosity on cell departure angle emphasizes the need to perform laboratory assays and ex vivo
sperm-tract interaction studies in medium with viscosity adjusted to the magnitude of physiological fluids. It also
suggests a possible role for viscosity in deflecting cells away from crypts in the reproductive tract. Our finding that
sperm respond to ratchet geometries in a similar way to bacteria may potentially improve microfluidic IVF devices,
through acting to direct high concentrations of motile cells towards the egg.
We only have the beginnings of an understanding of how the minute population of sperm reaching the site of fertilization may differ from the vast majority that do not. The existence of this distinguished subpopulation was suggested by in vivo studies in rabbits \cite{cohen1980} over 30 years ago, but the determinants of successful migration still remain mysterious; these characteristics may include motility, in addition to immunological markers and morphology.
Further experimentation may also enable development of a useful motility-based functional diagnostic or prognostic test for male fecundity. For example, observation of sperm in microchannels may reveal hitherto undiscovered swimming parameters underlying successful tract migration or navigation.

As shown above, sperm cell migration in a micro-channel crucially depends on the channel geometry. Cells swim along
boundaries and, if the two flat boundaries intersect, cells follow the corner. This has cardinal consequences for
modeling of the cell behavior. Instead of spreading through a 3-dimensional domain, many cells swim along 1-dimensional
folds. First, this entails that wall features such as ratchets can prescribe swimming direction. Second, the size of
the domain available to the swimmers is drastically reduced, so cells collide more often; this requires special
consideration when modeling the spreading of the entire population, either in microchannel environments or the female
reproductive tract. The increased likelihood of a sperm-sperm collision may also have a
more complex behavioral effect; when cells collide,
mechanotransduction may induce cell signaling, altering beat pattern
and hence migratory behavior.

The findings now indicate that recent advances in investigating sperm
chemoattractants not only need to take account of the rheology of the
fluid in which the cells are swimming \cite{kirkman2011}, but also the
three-dimensional architecture of the fluid domain.
The application of experimental and
computational fluid dynamics is beginning to reveal the complexity of
the system of sperm-tract interaction, one of the central unsolved
problems in reproductive science.

\section*{Materials and methods}
This study employed channels of a cross-section $100\times100$~$\mu$m to observe trajectories of individual freely
migrating human sperm in microchannels of basic geometrical configurations (corners, curves) and more complex features
(`ratchets'). Cell behavior in micro-channels of basic geometrical configurations was studied. Microchannels of
$100$~$\mu$m height were produced in elastomer (PDMS) by soft lithography \cite{xia1998} and then bonded to a glass
coverslip after oxygen plasma treatment. Swimming cells were observed through the glass wall of the channel using a CCD
camera equipped with a standard microscope objective. A green 100~mW diode laser equipped with a condenser was used as
the light source. For imaging of the whole channel, we utilized a 160~mm 2x objective attached with an extension tube
to a 4~Megapixel Basler avA2300-25gm camera run at 4 fps. Cell swimming was examined in fluid of three different
rheologies: 0\%, 0.5\%, and 1\% methylcellulose (M0512, Sigma-Aldrich, Poole, UK, approximate molecular weight 88,000)
was added to Earle's Balanced Salt Solution without phenol red, supplemented with 2.5~mM~Na pyruvate and 19 mM Na
lactate (06-2010-03-1B Biological Industries, Kibbutz Neuro Haemek, Israel), and 0.3\% w/v charcoal delipidated bovine
serum albumin (Sigma A7906). Semen samples were obtained by masturbation, at the Centre for Human Reproductive Science,
Birmingham Women's NHS Foundation Trust from normozoospermic research donors giving informed consent, after 2--4 days'
abstinence. Donors provided informed consent under Local Ethical Approval (South Birmingham LREC 2003/239). Experiments
were performed between 1 and 3 hours after the semen sample was produced. The raw semen was injected into the wide
`entry' branch of the channel from which cells naturally spread to the main section. Results shown are representative
of 5 donors.

Acquired images were processed in series of 200 to form superimposition images. Pixels at which the brightness
increased from frame $n$ to frame $n+1$ above a certain level were stained, so that only moving objects are visible.
Additionally, the image sequence was color-coded as $RGB$ order, i.e.\ cell positions in frames 1 and 2 are stained
red, frames 3 and 4 green, frames 5 and 6 blue, frames 7 and 8 red again and so on. Hence, the direction of cell motion
can be inferred from superposition images. One such image is shown in Fig.~\ref{sperm:image_channel_shapes}. Camera
resolution was 2.7~$\mu$m/pix, too coarse to resolve details of the cell head, but sufficient to determine its
position.

\section*{Acknowledgments}
DJS and JKB acknowledge funding from Birmingham Science City. PD acknowledges the award from Warwick Institute of
Advanced Study. The authors thank staff at Birmingham Women's Hospital and members of the Reproductive Biology and
Genetics Group, University of Birmingham, for assistance; the authors also thank Professor Howard Berg for comments on
the manuscript.


\begin{thebibliography}{10}

\bibitem{rothschild1963}
Rothschild.
\newblock 1963.
\newblock Non-random distribution of bull spermatozoa in a drop of sperm
  suspension.
\newblock {\em Nature}, 198:1221--1222.

\bibitem{winet1984}
H.~Winet, G.~S. Bernstein, and J.~Head.
\newblock 1984.
\newblock Observations on the response of human spermatozoa to gravity,
  boundaries and fluid shear.
\newblock {\em J. Reprod. Fert.}, 70:511--523.

\bibitem{cosson2003}
J.~Cosson, P.~Huitorel, and C.~Gagnon.
\newblock 2003.
\newblock How spermatozoa come to be confined to surfaces.
\newblock {\em Cell Motil. Cytoskel.}, 54:56--63.

\bibitem{woolley2003}
D.~M. Woolley.
\newblock 2003.
\newblock Motility of spermatozoa at surfaces.
\newblock {\em Reproduction}, 126:259--270.

\bibitem{ramia1993}
M.~Ramia, D.~L. Tullock, and N.~Phan-Thien.
\newblock 1993.
\newblock The role of hydrodynamic interaction in the locomotion of
  microorganisms.
\newblock {\em Biophys. J.}, 65:755--778.

\bibitem{fauci1995}
L.~J. Fauci and A.~McDonald.
\newblock 1995.
\newblock Sperm motility in the presence of boundaries.
\newblock {\em Bull. Math. Biol.}, 57:679--699.

\bibitem{smith2009jfm}
D.~J. Smith, E.~A. Gaffney, J.~R. Blake, and J.~C. Kirkman-Brown.
\newblock 2009.
\newblock Human sperm accumulation near surfaces: a simulation study.
\newblock {\em J. Fluid Mech.}, 621:220--236.

\bibitem{smith2009ms}
D.~J. Smith and J.~R. Blake.
\newblock 2009.
\newblock Surface accumulation of spermatozoa: A fluid dynamic phenomenon.
\newblock {\em The Mathematical Scientist}, 465:2417--2439.

\bibitem{blake1971}
J.~R. Blake.
\newblock 1971.
\newblock {A note on the image system for a Stokeslet in a no-slip boundary}.
\newblock {\em Proc. Camb. Phil. Soc.}, 70:303--310.

\bibitem{elgeti2010}
J.~Elgeti, U.~B. Kaupp, and G.~Gompper.
\newblock 2010.
\newblock Hydrodynamics of sperm cells near surfaces.
\newblock {\em Biophys. J.}, 99(4):1018--1026.

\bibitem{crowdy2010pre}
D.~G. Crowdy and Y.~Or.
\newblock 2010.
\newblock {Two-dimensional point singularity model of a low-Reynolds-number
  swimmer near a wall.}
\newblock {\em Phys. Rev. E}, 81(3 Pt 2):036313.

\bibitem{smith2011bj}
D.~J. Smith, E.~A. Gaffney, H.~Shum, H.~Gad\^{e}lha, and J.~Kirkman-Brown.
\newblock 2011.
\newblock {Comment on the article by J. Elgeti et al. Hydrodynamics of Sperm
  Cells Near Surfaces}.
\newblock {\em Biophys. J.}, 100:2318--2320.

\bibitem{elgeti2011}
J.~Elgeti, U.~B. Kaupp, and G.~Gompper.
\newblock 2011.
\newblock Response to comment on article: Hydrodynamics of sperm cells near
  surfaces.
\newblock {\em Biophys. J.}, 100(9):2321--2324.

\bibitem{lauga2006}
E.~Lauga, W.~R. DiLuzio, G.~M. Whitesides, and H.~A. Stone.
\newblock 2006.
\newblock {Swimming in circles: motion of bacteria near solid boundaries}.
\newblock {\em Biophys. J.}, 90(2):400--412.

\bibitem{li2009}
Guanglai Li and Jay~X. Tang.
\newblock Aug 2009.
\newblock Accumulation of microswimmers near a surface mediated by collision
  and rotational brownian motion.
\newblock {\em Phys. Rev. Lett.}, 103:078101.

\bibitem{giacche2010pre}
D.~Giacch{\'e}, T.~Ishikawa, and T.~Yamaguchi.
\newblock 2010.
\newblock {Hydrodynamic entrapment of bacteria swimming near a solid surface}.
\newblock {\em Phys. Rev. E}, 82(5):56309.

\bibitem{shum2010}
H.~Shum, E.~A. Gaffney, and D.~J. Smith.
\newblock 2010.
\newblock {Modelling bacterial behaviour close to a no-slip plane boundary: the
  influence of bacterial geometry}.
\newblock {\em Proc. R. Soc. Lond. A}.

\bibitem{berke2008}
A.~P. Berke, L.~Turner, H.~C. Berg, and E.~Lauga.
\newblock 2008.
\newblock {Hydrodynamic attraction of swimming microorganisms by surfaces}.
\newblock {\em Phys. Rev. Lett.}, 101(3):38102.

\bibitem{or2009}
Y.~Or and R.~M. Murray.
\newblock 2009.
\newblock {Dynamics and stability of a class of low Reynolds number swimmers
  near a wall}.
\newblock {\em Phys. Rev. E}, 79.

\bibitem{crowdy2010jfm}
D.~Crowdy and O.~Samson.
\newblock 2011.
\newblock Hydrodynamic bound states of a low-reynolds-number swimmer near a gap
  in a wall.
\newblock {\em J. Fluid Mech.}, 667:309--335.

\bibitem{suarez2006}
S.~S. Suarez and A.~A. Pacey.
\newblock 2006.
\newblock Sperm transport in the female reproductive tract.
\newblock {\em Hum. Reprod. Update}, 12:23--37.

\bibitem{hulme2008}
S.~E. Hulme, W.~R. DiLuzio, S.~S. Shevkoplyas, L.~Turner, M.~Mayer, H.~C. Berg,
  and G.~M. Whitesides.
\newblock 2008.
\newblock Using ratchets and sorters to fractionate motile cells of escherichia
  coli by length.
\newblock {\em Lab Chip}, 8(11):1888.

\bibitem{galajda2007}
P.~Galajda, J.~Keymer, P.~Chaikin, and R.~Austin.
\newblock 2007.
\newblock A wall of funnels concentrates swimming bacteria.
\newblock {\em J. Bacteriol.}, 189(23):8704.

\bibitem{binz2010}
M.~Binz, A.P. Lee, C.~Edwards, and D.V. Nicolau.
\newblock 2010.
\newblock Motility of bacteria in microfluidic structures.
\newblock {\em Microelectronic Engineering}, 87(5-8):810--813.

\bibitem{cohen1980}
J.~Cohen and K.R. Tyler.
\newblock 1980.
\newblock Sperm populations in the female genital tract of the rabbit.
\newblock {\em J. Reprod. Fert.}, 60(1):213.

\bibitem{kirkman2011}
J.C. Kirkman-Brown and D.J. Smith.
\newblock 2011.
\newblock Sperm motility: is viscosity fundamental to progress?
\newblock {\em Mol. Hum. Reprod.}, 17(8):539--544.

\bibitem{xia1998}
Y.~N. Xia and G.~M. Whitesides.
\newblock 1998.
\newblock Soft lithography.
\newblock {\em Annu. Rev. Mater. Sci.}, 28:153.

\end{thebibliography}

\end{document}